# Electrical transport crossover and large magnetoresistance in selenium deficient van der Waals HfSe$_{2-x}$ (0≤ $x$ ≤ 0.2)

Wenhao Liu[1], Yangzi Zheng[1], Aswin Lakshmi Narayanan Kondusamy[1], David L. Scherm[1], Anton. V. Malko[1], Bing Lv[1,2]*

1. Department of Physics, the University of Texas at Dallas, Richardson, Texas 75080, USA
2. Department of Materials Science & Engineering, the University of Texas at Dallas, Richardson, Texas, 75080 USA

Transition metal dichalcogenides (TMDs) have received much attention in the past decade not only due to the new fundamental physics, but also due to the emergent applications in these materials. Currently chalcogenide deficiencies in TMDs are commonly believed either during the high temperature growth procedure or in the nanofabrication process resulting significant changes of their reported physical properties in the literature. Here we perform a systematic study involving pristine stochiometric HfSe$_2$, Se deficient HfSe$_{1.9}$ and HfSe$_{1.8}$. Stochiometric HfSe$_2$ transport results show semiconducting behavior with a gap of 1.1eV. Annealing HfSe$_2$ under high vacuum at room temperature causes the Se loss resulting in HfSe$_{1.9}$, which shows unconventionally large magnetoresistivity following the extended Kohler's rule at low temperatures below 50 K. Moreover, a clear electrical resistivity crossover, mimicking the metal-insulator transition, is observed in the HfSe$_{1.9}$ single crystal. Further increasing the degree of deficiency in HfSe$_{1.8}$ results in complete metallic electrical transport at all temperatures down to 2K. Such a drastic difference in the transport behaviors of stoichiometric and Se-deficient HfSe$_2$ further emphasize the defect control and engineering could be an effective method that could be used to tailor the electronic structure of 2D materials, potentially unlock new states of matter, or even discover new materials.

**Introduction**

The large class of layered transition metal dichalcogenides (TMD) has emerged as a potential channel material alternative to silicon due to the requirements of ultrathin, high integrability, and low-power electronics for modern electronic systems. Due to their rich crystalline structures, wide variety of constituent elements, and the control offered via external perturbation such as chemical doping, proximity, gating, strain, and Moire patterning, many fascinating electronic and optical properties have been discovered in this family. For instance, strong correlation phenomena such as superconductivity charge density wave, heavy fermions, and Mott insulators are demonstrated [1–5]. 2D magnetism including room temperature FM, spirals, skyrmion type AFM, and possible quantum spin liquid are shown in TMDs doped with magnetic elements. Topology nontriviality, such as Weyl semimetals, is observed in T$_d$-WTe$_2$. [6–8]. In addition, coupling between ferroelectricity and superconductivity has been examined in T$_d$-WTe$_2$ and T$_d$-MoTe$_2$. Moreover, fascinating optical properties including strong photoluminescence and large optical excitation are demonstrated in WSe$_2$. [11,12]

In spite of this significant progress, many open questions and challenges remain. One major challenge is the difficulty of controlling and predicting the properties of materials with high deficiency or defects. Chalcogenide deficiencies are ubiquitous in TMD materials and can profoundly alter their mechanical, chemical, electrical, optical, thermal functionality, and their coupling with each other. Previous works show that defects not only modify the ground-state properties, excited-state properties, and a material's responses to external fields, but also could lead to new structures, unusual transport behaviors, novel magnetism, and even superconductivity [16,17]. To realize the full potential of the TMD system, understanding and controlling defects and deficiencies is required.

With this motivation, we completed one model study on the Se deficiency effect in HfSe$_2$. HfSe$_2$ adopts the 1T structure which consists of an octahedral prism and shows semiconducting behaviors with a bandgap of 1.13eV$^3$[18]. A high on/off current ratio exceeding 7.5 ×10$^6$ and high mobility is also demonstrated in HfSe$_2$, attracting considerable interest [19,20]. 1T-HfS$_2$ has ~2eV band gap while its sister compounds 1T-HfTe$_2$ shows metallic behaviors with a high magnetoresistance of 3000 % [9,10]. The bandgap in HfSe$_2$ can be tuned via external pressure or lithium intercalation[21,22], showing high potential for future electronic applications. HfSe$_2$ is not stable in air and previous work reveals the Hf metals in the surface reacted preferentially with oxygen, leading to the formation of more insulating HfO$_2$ islands or thin layer [23,24]. In this work, we find out the room-temperature storge under vacuum causes significant changes of the Se deficiency, and we further investigate the low-temperature transport behavior of HfSe$_{2-x}$ ($0 \leqslant x \leqslant 0.2$) through carefully controlling the Se deficiency. A clear transition from semiconducting behavior to metallic conductor behavior has been observed with the evolution of the Se deficiency. Moreover, unusually high magnetoresistance is observed in the low-temperature region (T < 25 K) for the HfSe$_{1.9}$ sample. This sample also follows the extended Kohler's rule in the low-temperatures below 50 K.

**Experiment Details:**

Single crystals were synthesized using a two-step process. First, polycrystalline samples are synthesized using Hf pieces (99.8%), Se shots (99.999%) in appropriate ratios (discussed below). The samples are firstly heated at 600 °C for 3 days in an evacuated silica tube followed by furnace cooling to get the polycrystalline precursors. In the second step, single crystals are synthesized using the chemical vapor transport (CVT) method. The pre-formed polycrystalline samples are sealed in evacuated silica tubes with I$_2$ as the transport agent (1 mg/cm$^3$). Plate-like large single crystals with dimensions of 3mm*3 mm are obtained with a two-week reaction time and source and growth zone temperatures fixed at 950 °C and 850 °C, respectively. 5% extra Se (i.e Hf:Se= 1: 2.1) is needed to obtain the stoichiometric HfSe$_2$ crystals. HfSe$_{1.8}$ crystals can be obtained through CVT synthesis with a Hf: Se ratio larger than 1: 1.7. The optimized condition to obtain HfSe$_{1.8}$ crystals with uniform Se deficiencies are with the ratio of Hf: Se=1:1.5 with the temperature profile of 950 °C (source) and 850 °C (sink) for two weeks.

The control of the HfSe$_2$ and HfSe$_{1.8}$ stoichiometry could be done easily by tuning the Hf and Se ratio in the CVT process, however, the precise control of the Se deficiencies down to 0.1 in the CVT-grown process is rather difficult. These reactions often yield batches of crystals with nonuniform Se deficiencies. Precise growth control of HfSe$_{1.9}$ single crystals could be obtained through post-annealing the as-grown stoichiometric HfSe$_2$ crystals in a silica tube under ultrahigh vacuum at 350 °C for 1-5 days. The best control of HfSe$_{1.9}$ samples can be obtained via annealing HfSe$_2$ at room temperature under high vacuum or in the glovebox with inert atmosphere for three months. Carefully comparing the annealed samples with the as low-yield as-grown HfSe$_{1.9}$ has been done to ensure our sample quality and our observation are intrinsic. This room temperature annealing effect on HfSe$_2$ also highlights the essential need for proper storage of the samples.

The exact chemical composition of the crystals obtained was verified by energy-dispersive X-ray spectroscopy (EDX) on a DM07 Zeiss Supra 40 scanning electron microscopes. X-ray diffraction was conducted on single crystal samples using a Rigaku Smart Lab X-ray diffractometer equipped with Cu-Kα radiation. Resistivity was conducted on the freshly cleaved surfaces of HfSe$_{2-x}$ using the four-probe method in Quantum design Physical Property Measurement System (PPMS) down to 1.8 K.

The single-crystal X-ray data were measured on a Bruker SMAER diffractometer with an Apen II area detector with a Mo Kα source ($\lambda = 0.71073$ Å).

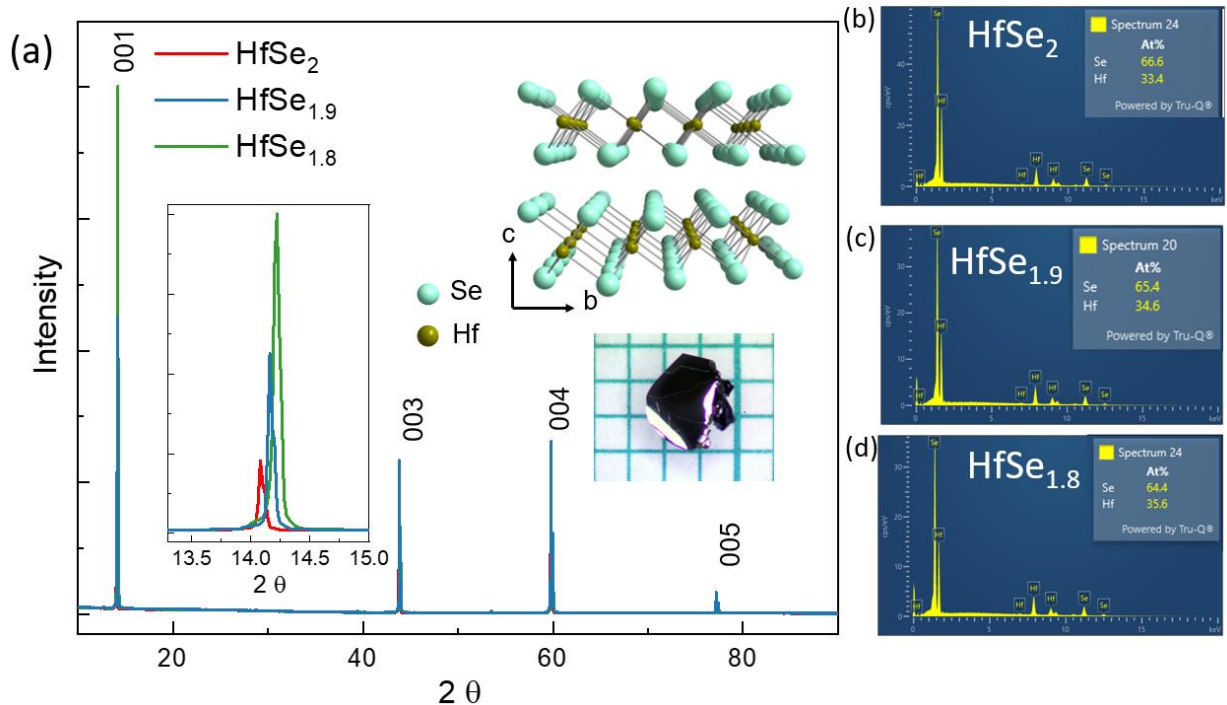

Figure 1. (a) X-ray diffraction pattern of $HfSe_{2-x}$ single crystals. Left inset: enlarged view of (*001*) peak of $HfSe_{2-x}$ single crystals. Right top inset: illustration of the crystal structure of $HfSe_{2-x}$. Right bottom inset: optical image of as-grown $HfSe_2$ single crystal. (b)-(d) EDX spectra of $HfSe_{2-x}$ upon increasing Se deficiency.

Fig. 1a shows the X-ray diffraction (XRD) patterns of $HfSe_{2-x}$ crystals with different deficiencies. The inset shows the optical image of $HfSe_2$ on a millimeter scale grid with a freshly cleaved surface used for X-ray diffraction. All $HfSe_{2-x}$ crystalize in the 1T structure. Only the *00l* peaks are shown, demonstrating the crystallographic *c* axis is perpendicular to the flat surface of the single crystal. The bottom-left inset shows that upon increasing Se deficiencies, the X-ray peaks shift slightly towards higher angles, indicating the small reduction of the out-of-plane lattice parameter with the increasing Se deficiency. The refined *c* lattice parameters for $HfSe_2$, $HfSe_{1.9}$ and $HfSe_{1.8}$ are *c* =6.169(2), 6.156(7), 6.139(1) Å, respectively. Notably, there is no additional XRD peak observed beyond the *(00l)* peak with increasing Se deficiency. Furthermore, to rigorously validate the crystal structure, a single crystal X-ray diffractometer was employed. All specimens from three batches of $HfSe_2$, $HfSe_{1.9}$ and $HfSe_{1.8}$ consistently reveal identical structures, affirming the absence of any secondary phases. The element ratio is confirmed by energy-dispersive X-ray spectroscopy (EDX) in Fig.1b-d, which reveals the Hf: Se molar ratio as 1:1.97±0.03, 1:1.88±0.01 and 1:1.79±0.03. For convenience, we use $HfSe_{1.8}$, $HfSe_{1.9}$ and $HfSe_2$ to indicate these phases, respectively. Such small standard deviations indicate high homogeneity of the element's distribution, and high quality of our samples.

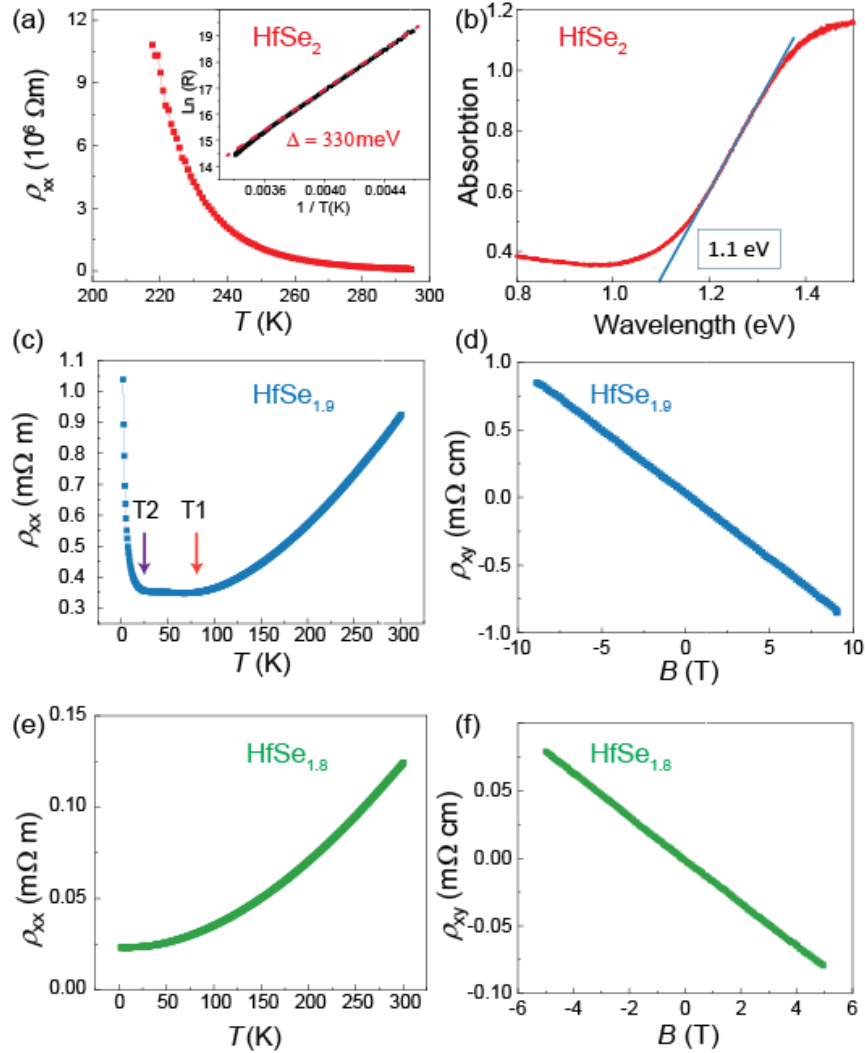

Figure 2. (a) Temperature dependence of resistivity in stoichiometric HfSe$_2$. Inset: Ln (R) vs. 1/T and the thermal activation gap. Red dashed Line: fitting using thermal activation model. (b) Optical absorption spectra of stoichiometric HfSe$_2$. (c) (d) Temperature dependence of resistivity and Hall resistivity of HfSe$_{1.9}$. (e) (f) Temperature dependence of resistivity and Hall resistivity of HfSe$_{1.8}$.

The temperature-dependent electrical resistivity data for three crystals are shown in Fig. 2. The resistivity of stoichiometric HfSe$_2$, as indicated in Fig. 2 (a), increases upon cooling of the temperature, demonstrating typical semiconducting behavior. The sample exceeds the upper resistance limit of our equipment below 200 K. The resistivity can be fit quite well using the thermal activation model $\rho = \rho_0 exp (E_a/K_B T)$, where $\rho_0$ is a prefactor and $k_B$ is the Boltzmann constant. The Fig. 2a inset shows the results of linear fitting of ln($\rho$) vs. (1/$T$), where the activation energy is estimated to be ~330 meV.

Fig.2(b) shows the optical absorption spectra of stoichiometric HfSe$_2$. To properly measure absorption spectrum of HfSe$_2$ crystal, we employed microscope-based system which can measure optical absorption spectra from micron-sized samples. The HfSe$_2$ flake and a reference sample were illuminated by a fiber-coupled white light source (Thorlabs SLS201L) through a 20X objective lens. The transmission spectra collected by the microscope passed through a monochromator and were recorded by a CCD camera. By applying the Tauc method, the estimated optical band gap of HfSe$_2$ was found to be about 1.1 eV.

Fig. 2(c) shows the temperature dependence of resistivity of HfSe$_{1.9}$ where a clear resistivity crossover is observed. From room temperature temperatures higher than 100 K (indicated by T1), dR/dT is positive, indicating metallic behavior in this range. Between 20 K and 100 K, resistivity stays nearly constant, and starts to rise again when the temperature is below 20 K (indicated by T2). The resistivity values at room temperature (0.95 mΩ m) are fairly comparable to that at 2 K (1.05 mΩ m) but are significantly smaller than the room temperature value (8000 mΩ m) for the HfSe$_2$ sample, suggesting the effect of charge carriers increasement is more dominant than the impurity scattering effects caused by defects. The Hall resistivity of HfSe$_{1.9,}$ shown in the inset of Fig.2(c), exhibits a linear relation with external magnetic fields. The slope is negative, demonstrating charge carriers are mainly electrons and the estimated carrier density from Hall data using single band model is about $6.6 \times 10^{18}$ cm$^{-3}$. Further increasing the amount of Se deficiency to produce HfSe$_{1.8}$ leads to the complete metallic behavior across the whole temperature range as illustrated in Fig.2(d). Further Hall effect measurements suggest the further increase of the electron charge carriers to $3.9 \times 10^{19}$ cm$^{-3}$, accounting for the emergence of complete metallic behavior for HfSe$_{1.8}$ phase.

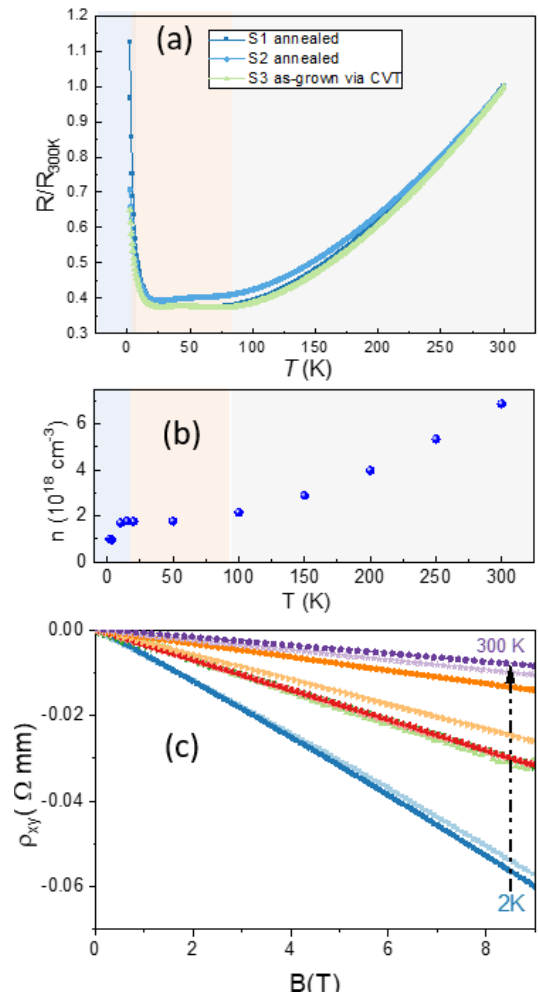

Figure 3. (a) Temperature dependence of normalized resistance of annealed HfSe$_{1.9}$. S1 and S2 indicate two samples from different batches synthesized via annealing and S3 indicate low yielding as-grown HfSe$_{1.9}$ from synthesized via CVT. (b) Temperature-dependent carrier concentration of HfSe$_2$ determined from the Hall resistivity in (c).

To further understand the metal-insulator-like resistivity crossover in the HfSe$_{1.9}$ sample, low-temperature Hall measurements at various temperatures were carried out. Overall linear dispersion with negative slopes were measured in the whole temperature range investigated, reinforcing that the carrier density is dominantly electrons near the Fermi surface. A comparison between electrical resistivity and calculated carrier density from Hall data at different temperatures is plotted in Fig. 3. The carrier density decreased monotonically with decreasing of the temperature. This nicely coincides with the corresponding temperature-dependent resistivity results, suggesting the change of electron charge carriers is the main reason for the electrical resistivity crossover at different temperatures. For temperatures below 20 K, the carrier density decreased below $1\times10^{18}$ cm$^{-3}$, and this low carrier density may cause the insulating behavior in the low temperature region.

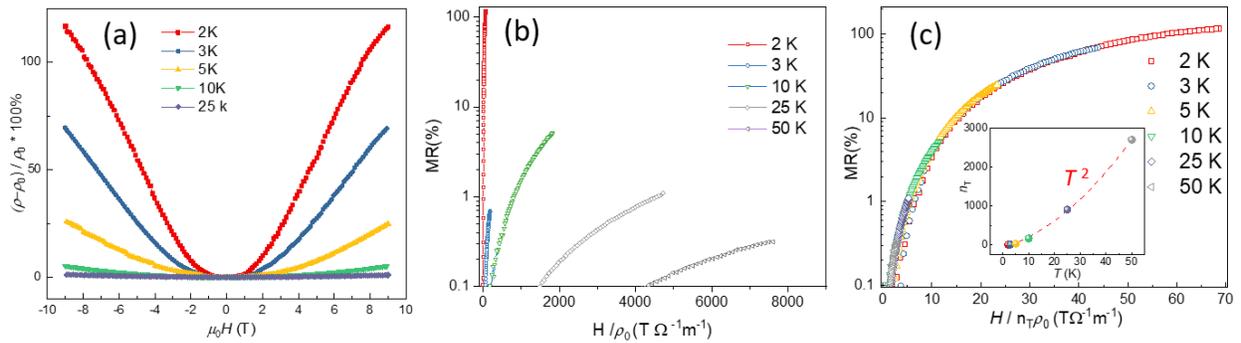

Figure 4 (a) Magnetoresistance of HfSe$_{1.9}$ up to 9T at various temperatures. (b) The violation of Kohler's rule for the MR in (a). (c) Test for extended Kohler's scaling of magnetoresistance in HfSe$_{1.9}$ single crystal.

Figure 4(a) shows the magnetoresistance of HfSe$_{1.9}$ up to 9 T at different temperatures. At 2 K, the MR evolves gradually from a positive curvature to a negative curvature when the external magnetic field was increased above 6 T. This does not follow the common quadratic behavior. Magnetoresistance ratio, which is defined as $MRR = (\rho(B)/\rho_0 - 1) \times 100\%$, can reach 120 % at 2 K with an external magnetic field of 9 T. This value is quite high considering the absence of magnetic elements in this material.

Large magnetoresistance is often linked with topologically non-trivial band structure. Prime examples of this are WTe$_2$ and MoTe$_2$ [25,26]. In Hf-based compounds such as HfTe$_2$, large nonsaturating magnetoresistance has been observed [10,27], which was attributed to carrier compensation as initially HfTe$_2$ is reported to be trivial semimetal with coexisting electrons and holes at the Fermi surface. But recently some clear Dirac-like cone features at the center of the Brillouin zone are observed from ARPES studies on the high-quality MBE-grown monolayer thin films, suggesting materials quality might be crucial for further studies of this system. Our HfSe$_{1.9}$ sample here likely undergoes very subtle electronic structure changes compared to the pristine HfSe$_2$ sample. Meanwhile, the more Se deficient HfSe$_{1.8}$ sample shows the most encountered quadratic behavior with a MRR of about 1%, demonstrating a normal metallic phase in the compounds with more Se deficient.

In semiclassical transport theory, the temperature and magnetic dependence of resistance can often be analyzed via Kohler's rule[a], which dictated that the magnetoresistance MR obeys the scaling behavior of $f[H/(\rho_0)]$ where MR = $[\rho(H)-\rho_0]/\rho_0$. Here $H$ is the magnetic field, with $\rho(H)$ and $\rho_0$ being the resistivity at H and zero field, respectively. Kohler's rule holds if there is a constant single carrier type and the scattering time τ is the same on all points on the Fermi surface. [25, 28], as demonstrated in various metals. [29] The validity of Kohler's rule has been extended to several semiconductors and even in cuprate superconductors if the Fermi surface remains largely temperature independent [30-32]

In $HfSe_{1.9}$, MR curves at fixed temperatures vs. H/ $\rho_0$ do not collapse into one single curve, demonstrated the violation of Kohler's rule, as illustrated in Fig.4(b). The violation of Kohler's rule can be caused by different mechanisms including multiple scattering rates, [33-35], multiband effect [36, 37], carrier density change induced by Fermi surface shift or temperature [38,39] etc. which usually implies the possibility of new emergent physics phenomena hidden in this phase.

Recently, several other models have been proposed for the case where Kohler's rule is violated. For example, the extended Kohler's rule MR= $f[H/(n_T\rho_0)]$ is raised and accounts for the systems with change of total carrier density. [40]. Here $n_T$ is responsible for the carrier change. In $HfSe_{1.9}$, a clear temperature dependence of carrier density can be observed, as shown in Fig. 3(c). The unusual large magnetoresistance collapse into one single line if $n_T$ is considered in H/ $\rho_0$, following of extended Kohler's rule, as shown in Figure 4(c). Here we define $n_T$ at 2 K equal to 1 and temperature dependence of $n_T$ is shown in the inset of Figure 4(c). $n_T$ obtained in our sample follows a rough $T^2$ scaling, which can be attributed to the thermal induced change in the carrier density, as can be seen in other systems like TaP [40]

In conclusion, we reported the room temperature effect on the chemical stoichiometry and carried out systematic research on the impact the defects play on a model system $HfSe_{2-x}$. Stoichiometric $HfSe_2$ shows semiconducting behavior with a band gap of 1.1 eV as determined by the optical absorption spectrum. Se deficiency causes electron doping in the system, which will tune the carrier density and the fermi level, changing the system from semiconducting $HfSe_2$ to metallic $HfSe_{1.8}$. In the metastable middle compound $HfSe_{1.9}$, competition between the metallic and insulating phases becomes more paramount and may account for the large magnetoresistance and following the extended Kohler's scaling. The rich phase transitions of $HfSe_{2-x}$ provide another material platform to investigate the mechanism of Kohler and extended Kohler's scaling. Besides, the differences between stoichiometric and room temperature annealed $HfSe_2$ further demonstrates that the preparation, storage time and defect/level should be clarified precisely in the chemical and physical property analysis, especially the device characterization of the TMD family.

## Acknowledgements

This work at the University of Texas at Dallas is supported by US Air Force Office of Scientific Research Grant No. FA9550-19-1-0037, National Science Foundation (NSF)- DMREF- 2324033 and Office of Naval Research (ONR) grant no. N00014-23-1-2020. Our measurement facilities also acknowledge the support from the AOSFR Defense University Research Instrumentation Program (DURIP) grant no. FA9550-21-1-0297.